\documentclass{article}

\usepackage{arxiv}

\usepackage{amsmath}
\usepackage{amsfonts}
\usepackage{amssymb}
\usepackage{comment}
\usepackage{tabularx}
\usepackage{url}
\usepackage{subcaption}
\usepackage{graphicx}

\usepackage{biblatex}
\title{Visualization of Knowledge Graphs with Embeddings: an Essay on Recent Trends and Methods}

\author{{Davide Riva} \\
	Department of Computer Science, Università degli Studi di Milano\\
        Milano, Italy \\
	Department of Control and Computer Engineering, Politecnico di Torino\\
	  Torino, Italy \\
	\texttt{davide.riva1@unimi.it} \\
	\And
	{Cristina Rossetti} \\
	Department of Information Engineering, Università Politecnica delle Marche\\
	Ancona, Italy\\
	Department of Control and Computer Engineering, Politecnico di Torino\\
	  Torino, Italy \\
	\texttt{cristina.rossetti@polito.it}
    }

\usepackage{hyperref}
\let\svthefootnote\thefootnote
\newcommand\freefootnote[1]{%
  \let\thefootnote\relax%
  \footnotetext{#1}%
  \let\thefootnote\svthefootnote%
}

\begin{document}

\flushbottom
\maketitle
\thispagestyle{empty}

\freefootnote{The authors contributed equally to this work.}

\begin{abstract}
In this essay we discuss the recent trends in visual analysis and exploration of Knowledge Graphs, particularly in conjunction with Knowledge Graph Embedding techniques. We present an overview of the current state of visualization techniques and frameworks for KGs, in relation to four identified challenges. The challenges in visualizing Knowledge Graphs include the need for intuitive and modular interfaces, performance in handling big data, and difficulties for users in understanding and using  query languages. We find frameworks that generally satisfy the intuitive UI, performance, and query support requirements, but few satisfying the modularity requirement. In the context of Knowledge Graph Embeddings, we divide the approaches that use embeddings to facilitate exploration of Knowledge Graphs from those that aim at the explanation of the embeddings themselves. We find significant differences between the two perspectives. Finally, we highlight some possible directions for future work, including diffusion of the unmet requirements, implementation of new visual features, and experimentation with relation visualization as a peculiar element of Knowledge Graphs.
\end{abstract}

\section{Introduction}
In the last decade, the Semantic Web has emerged as a global interconnected data space \cite{linkeddata} that facilitates the integration and the exploration of heterogeneous data. In this context, Knowledge Graphs (KGs) have been developed as a graph-based semantic technology that aims to represent the semantics of information of a certain domain in a flexible and structured way.
Nowadays, KGs are used in various fields of application, in academic as well as industrial context. Given that a KG is an enabling technology to semantically represent and integrate heterogeneous data from different sources, an user may be interested in exploring and visualizing stored information. Here Data Visualization comes into play as an important step in data analytics, as it allows to better understand data, find hidden patterns or display data from different points of view. This process can be particularly useful when data exploration and analysis objectives are not well defined (\cite{VizKG}). Research on data visualization, in an effort to address these problems, aims to find the best solutions to create intuitive, interactive and scalable interfaces based on different purposes. \\ 
In the context of KGs, Data Visualization supports and facilitates the navigation of the information stored in the graph, handling its complexity and overcoming the need for expertise in query languages. In recent times, Graph Embedding technologies have also been exploited to address the task of retrieving information from a KG (\cite{hamilton2018embedding}), hence a conspicuous line of research has focused on how visual exploration of KGs might benefit from them. Indeed, Knowledge Graph Embedding (KGE) techniques and the visualization of their output have been used in a variety of fields, ranging from hardware security (\cite{utyamishev2022knowledge}) to biomedical cause-effect relation discovery (\cite{kramer2022mining}). At the same time, KGE technologies are often based on black-box algorithms that hinder transparency and interpretability of their output. Besides benefiting from graph embeddings, visualization can also support our understanding by enabling qualitative analysis and comparison. \\
This work aims at studying the current state of the art of the visualization techniques for Knowledge Graphs, with a special attention dedicated to the integration with Graph Embedding techniques. To the best of our knowledge, no survey has been produced so far concerning this perspective. \\
In the following, we provide first an overview of KG and KGE technologies (Section \ref{sec:background}) and the challenges that arise (Section \ref{sec:rel-work}). Then we define our survey methodology (Section\ref{sec:methodology}) and we review the state of the art frameworks in the visualization of KGs in general (Section \ref{sec:KG_visualization}) and in conjunction with KGEs (Section \ref{sec:KGE_visualization}). We draw our conclusions in Section \ref{sec:concl}.

\section{Background}\label{sec:background}

KGs have seen a recent surge in popularity in Computer Science and Artificial Intelligence thanks to their expressive power and the variety of applications they can serve. The diffusion of new graph embedding techniques, specifically tailored to KGs, provided new opportunities for querying the graphs as well as for their application in AI. In this section we give an overview of the state-of-the-art of these technologies.

\subsection{Knowledge Graphs and Applications}
A Knowledge Graph is a semantic technology born with the expansion of the Semantic Web and first introduced by Google in 2012. 
In order to give a formal definition, we refer to the formulation of \cite{kgdefinition}: \\
\textit{"A Knowledge Graph (i) mainly describes real world entities and their interrelations, organized in a graph, (ii) defines possible classes and relations of entities
in a schema, (iii) allows for potentially interrelating arbitrary entities with each other, and (iv) covers various topical domains"}.\\
With reference to point (i) and (iv), a Knowledge Graph is a base of knowledge of a certain domain, defining a graph-based global access schema that semantically describes in a formal and unambiguous way entities and their relations. According to point (ii), the domain of interest is formally represented with classes, i.e sets of entities, and relations. With reference to the theory of Knowledge Bases (KBs), a KG is also a special kind of KB whose relations (also known as \textit{predicates}) are of arity $n \leq 2$, i.e. it involves only unary and binary relations, and therefore can be conveniently represented as a graph. \\
A KG can be created with different implementations. The most common format for KG representation is the Resource Description Format (RDF), which uses triples (subject, predicate, object) to build entities, classes and their relations in a graph-based data model. With RDF graphs, the standard query language is SPARQL. \\
Ontologies are another semantic technology strictly related to KGs. According to \cite{ontologydefinition}, the term \textit{ontology}, in this context, refers to a set of representation primitives (e.g. classes, attributes and relations) for structuring the knowledge of a certain domain. Primitives may also include a semantic description and constraints on data. The most widely acknowledged reference standard is the Web Ontology Language (OWL) and RDF-Schema (RDFS), both useful for modeling RDF graphs, but many other examples exist, such as DBPedia-Ontology (DBO) or the Simple Knowledge Organization System (SKOS). So the formal definitions given by an ontology can be useful to build a Knowledge Graph with reference to a common vocabulary. \\
\newline
In last years, Knowledge Graphs have gained a lot of attention and many applications have been proposed. According to \cite{kg2024}, KGs are now ``\textit{an industry standard for data unification, question-answering, recommender systems, explainable AI}''. Indeed, Knowledge Graphs are widely used for data integration and to structure Big Data management systems like Data Lakes. Hassanzadeh et al. (\cite{DMSW}) argued that the introduction of semantic technologies like KGs in a data management system allows a faster integration process, the generation of mappings between concepts, a better data sources search and identification. 
Semantic Data Lakes, for instance, support the query over data sources stored in a Data Lake by representing them in a KG and defining a set of logical rules to reason over them (\cite{adbis2022,oro83013,aurum}) \\
Another use of KGs is in \textit{recommender systems}: automatic systems that search for relevant items with the aim to provide recommendations based on users' interest (e.g. online advertising). Here KGs are navigated starting from user's interests (\cite{ripplenet}) or fed into Graph Neural Networks (\cite{KGCN}) to produce recommendations. 
Furthermore, many works have been published about the use of KGs in medicine and healthcare (\cite{abu2023healthcare,medical_KG1,medical_KG2}),  cybersecurity (\cite{KG_cyber1,KG_cyber2}), Industry 5.0 (\cite{arazzi2024semioe}), geoscience (\cite{KG_geoscience}), education (\cite{KG_ed}), and Natural Language Processing (\cite{zhang2021greaselm}).

\subsection{Knowledge Graphs Embeddings}
\label{KG_embeddings}

\begin{figure}
    \centering
    \includegraphics[width=0.8\textwidth]{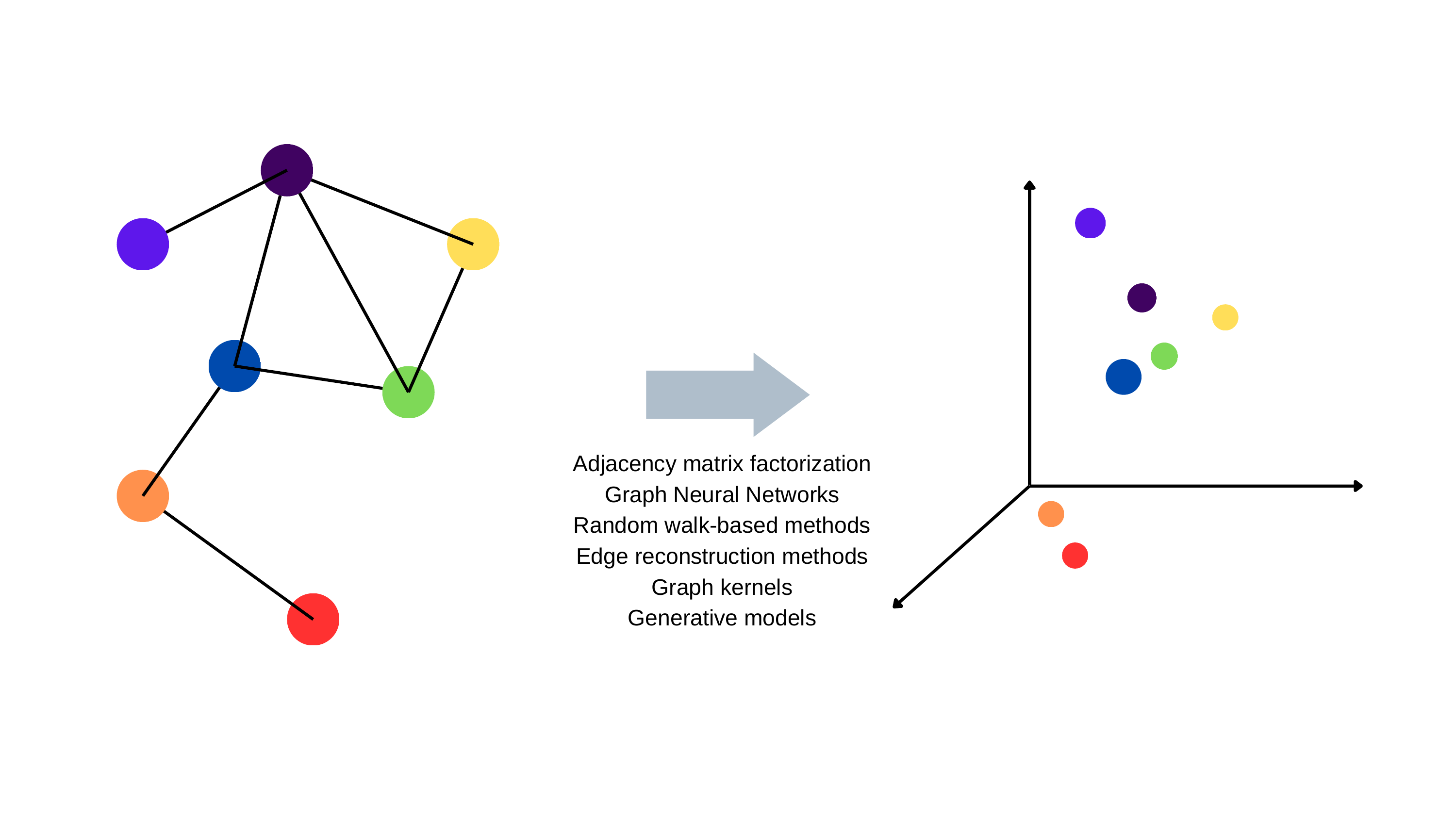}
    \caption{Schema of graph embedding.}
    \label{fig:ge}
\end{figure}

Graph Embedding models are mathematical functions that map the nodes of a graph $G = (N,E)$ onto a vector space $\mathbb{R}^d$, whose dimension $d$ is typically in the order of $10^2$ or $10^3$. These methods rely on a wide variety of machine learning techniques, with the common goal of representing similar nodes, i.e. nodes that share the same links and/or attributes, as close vectors in the embedding space, as displayed in Figure \ref{fig:ge}. For general graphs they rely on a wide variety of techniques, usually categorized into: matrix factorizations \cite{cao2015grarep}, random walk-based methods \cite{grover2016node2vec}, Graph Neural Networks (GNNs)\cite{fu2020magnn}, edge reconstruction methods \cite{bordes2013translating}, graph kernels \cite{yanardag2015deep}, and generative methods \cite{xiao2017ssp}. In the case of KGs, the most widely adopted methods are edge reconstruction and GNN-based methods.\\
Such models can be applied to KGs to represent their semantics in a relatively low-dimensional vector space. This enables to perform tasks such as \textit{link prediction}, which aims at predicting the object of a triple given the subject and the predicate, \textit{relation classification}, which predicts the predicate given subject and object, or \textit{community detection}, which clusters entities that have common properties, by operating with vectors.\\
KG Embedding techniques serve both the purpose of querying KGs and that of injecting information from KGs into AI systems.  \textcite{huang2019knowledge} uses entity and relation embeddings to perform approximate queries, i.e. vector similarity queries, over a KG and answer questions. Other applications include the injection of background knowledge into a Language Model (see, for instance, \textcite{Peters2019KnowledgeEC}) via interaction between word embeddings and KG embeddings.\\
One limitation of this technology is that the embeddings of entities and relations often lack interpretability, as they are produced by black-box models. Therefore, visualization tools can not only benefit from KGE models but also qualitatively support the analysis of their output so to assess the information captured in the vector space.

\section{Challenges in Visualization}\label{sec:rel-work}
The visualization of a Knowledge Graph can pose several challenges. First of all, an effective visualization must be intuitive and clear so the user can easily explore and analyse the represented information. Furthermore, nowadays the amount of information stored and needed may be huge, so KGs can be very large. As as example, DBPedia \cite{auer2007dbpedia} currently contains a number of entities in the order of $10^8$ and a number of triples in the order of $10^{10}$. Displaying a large graph in a limited space may be difficult as it creates problem like overlaps of nodes and edges (\cite{kgvisualization2020}), leading to a poor understanding of the context and the representation in general. Big Data management in a Knowledge Graph can also lead to challenges in the performance of the visualization framework. Indeed, the visualization process may cause decreases in responsiveness and may put an excessive burden on low-performance machines (\cite{kgvisualization2020}).\\
Moreover, user can find difficulties in understanding and learning query languages (\cite{kg2024}) for graphs, like SPARQL in RDF graphs. So frameworks for KG visualization should consider interfaces provided with functions that facilitate the exploration and query of the graph. With a high level of abstraction from the language, the accessibility of the framework, and consequently of the KG, increases as also non-expert users can easily utilize it. \\ 
Challenges may also refer in general to the way in which the user visualizes and focuses on details of the KG. This can be a problem for several reasons: (i) users may find it difficult to understand the context as KGs can describe a large variety of concepts in a certain domain, (ii) it is possible that non-expert end-users do not have the capabilities to handle graph's structures, (iii) analysing data and get useful insights from a graph visualization may not be easy so frameworks should help users to effectively retrieve crucial information. \\
Moreover, frameworks should be adaptive to different kinds of representations and query languages, providing a general-purpose architecture. The visualization approaches should also consider to give the other developers the possibility to extend or modify the base architecture with additional plugins or other functionalities based on needs. \\
Summing it up, the main goals that a visualization framework for KGs has to achieve are the following: 
\begin{itemize}
    \item  \textit{Modularity}:  the system has to be well-structured, adaptable and flexible to changes and additional functionalities. It also includes the possibility to be adaptive to any kind of graph structure and language as well as to domains of application;
    \item  \textit{Intuitive UI}: it means that (i) the user interface has to be clear and simple to use, (ii) the visualization should not rely much on the KG's structure, (iii) the system should facilitate as much as possible the understanding of results and the finding of valuable insights;
    \item \textit{Performance}: how much the framework is capable to handle Big Data in KG representation;
    \item \textit{Query Support}: the system should offer functionalities to simplify the exploration, interaction and analysis of a KG by implementing abstraction, which relieves the user from learning and using a KG query language.
\end{itemize} 

We acknowledge that the evaluation of some of these aspects in existing visualization frameworks is naturally subject to personal perspective. However, we try to keep an objective eye by inspecting the features that are present and those that are absent in each framework. For instance, for what concerns the \textit{Intuitive UI} requirement, we focus mainly on point (ii), i.e. independence from KG structure, which appears to be the most objective factor.

\section{Survey Methodology} \label{sec:methodology}

This survey covers literature from the year 2014 up to September 2024 concerning publications on KG visualization frameworks. The research was conducted on Scopus with the following search query: \texttt{TITLE-ABS-KEY ( "knowledge graph" ) AND TITLE-ABS-KEY ( "visualization" OR "data visualization" OR "visual analytics" OR "visual analysis" )}. Only publications in English language, within the computer science field (Subject Area filter) and on conference proceedings and journals were selected. A total of 656 documents was retrieved from this research, 437 of which were on conference proceedings while 219 on journals. Of all these papers, 72 were included in this survey, while the others were excluded for several reasons, including:
\begin{itemize}
    \item Low relevance to the scope of the survey, i.e. frameworks or tools not tailored to KG visualization;
    \item Works on data visualization of KGs using the same visualization tool;
    \item Closed-access papers;
    \item Low-ranking journals or conferences.
\end{itemize}

\begin{figure}[ht]
	\centering
    \includegraphics[angle=0,origin=c,width=0.5\textwidth]{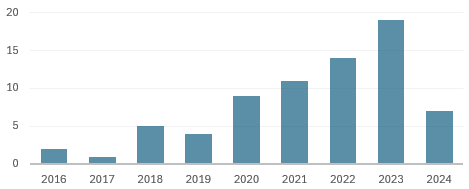}
    \caption{Number of papers grouped by year of publication}
    \label{fig:p_per_year}
\end{figure}

The distribution of selected publications over years is shown in Figure \ref{fig:p_per_year}. The trend displayed in the chart shows a growing interest in the field, which also underpins the motivation for a comprehensive survey.

\section{Knowledge Graphs Visualization}
\label{sec:KG_visualization}
A survey of Knowledge Graphs visualization techniques is given in this section, considering the challenges faced, the main features of the different frameworks and their accomplished goals. In table \ref{table:kgvisualization} frameworks, sorted by year of publication, are summarized with their achieved goals. In the following, visual examples are also provided for selected (and accessible) frameworks.

\begin{table}[ht]
    \begin{tabularx}{\textwidth} { 
  | >{\raggedright\arraybackslash}X 
  | >{\centering\arraybackslash}X 
  | >{\centering\arraybackslash}X |
    >{\centering\arraybackslash}X |
    >{\centering\arraybackslash}X |}
    \hline
    \textbf{Visualization framework} & \textbf{Modularity} & \textbf{Intuitive UI} & \textbf{Performance} & \textbf{Query Support}\\
    \hline
     AllegroGraph \cite{AllegroGraph} & & \checkmark & \checkmark & \checkmark \\
    \hline
    StarDog Explorer \cite{StarDog} & & \checkmark & \checkmark & \checkmark \\
     \hline
     LodLive \cite{LodLive} & \checkmark & \checkmark  &  & \checkmark \\
    \hline
    TigerGraph GraphStudio \cite{TigerGraph} & \checkmark & \checkmark & \checkmark & \checkmark \\
     \hline
    EDUVIS \cite{sun2016eduvis} &  & \checkmark &  &  \\ 
    \hline
     ALOHA \cite{ALOHA} &  & \checkmark & & \checkmark \\
    \hline
     CEPV \cite{CEPV} &  &  & \checkmark & \checkmark \\
    \hline
     Nararatwong et al. \cite{kgvisualization2020} & \checkmark & \checkmark & \checkmark & \checkmark \\ 
    \hline
     VizKG \cite{VizKG} & \checkmark & \checkmark &  & \checkmark \\
     \hline
     Lim et al. \cite{KGcovid} & & & & \\
    \hline
     KGNav \cite{wang2023kgnav} &  & \checkmark & & \checkmark \\
     \hline
  \end{tabularx}
  \caption{Summary of KGs visualization techniques, ordered by publication year.}
  \label{table:kgvisualization}
\end{table}

AllegroGraph \cite{AllegroGraph} is a closed-source software developed to store and manage RDF triples. Up to now, the developers have provided different products and functionalities like Neuro-Symbolic AI capabilities or Cloud platform. One of the products, called Gruff, is KG-specific and it augments KG exploration with intuitive visualizations. Examples of available tools are the highlighting of relevant relationships or the discovery of hidden patterns. A visual query builder is provided to help non-expert users to create highly complex queries without any previous knowledge of query languages. This specific module is open-source and also available on browser with an underlying cloud technology allowing high-performance executions. Many functionalities are offered in Gruff, such as the possibility to filter nodes, to change the visualization by modifying colors and sizes, or to add images to nodes.

\begin{figure}[ht]
	\centering
    \includegraphics[angle=0,origin=c,width=0.5\textwidth]{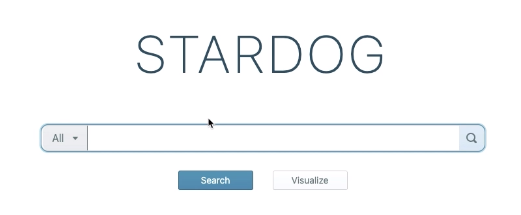}
    \caption{Example of entities search from StarDog Explorer\cite{StarDog}}
    \label{fig:stardog_browser}
\end{figure}

\begin{figure}[ht]
	\centering
    \includegraphics[angle=0,origin=c,width=0.9\textwidth]{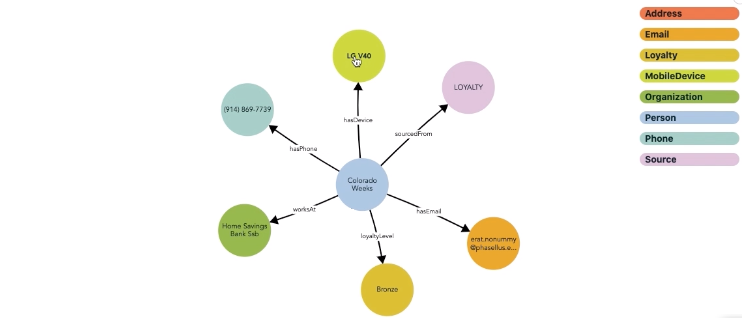}
    \caption{Example of graph visualization from StarDog Explorer \cite{StarDog}}
    \label{fig:stardog_graph}
\end{figure}

StarDog Explorer \cite{StarDog} is a product created by StarDog Union company for KGs visualization and exploration. The framework provides a visual query support allowing users to create advanced queries without the need to have specialized knowledge about graph query languages. Furthermore the interface is intuitive and KG-agnostic because the exploration may also start with a browser-like search bar in which the user can insert the desired information (Figure \ref{fig:stardog_browser}). Search results are first shown as a list of entities related to the search that the user can expand and explore to better understand the context. Then, the displayed graph (Figure \ref{fig:stardog_graph}) is provided with different colors and shapes but also with clear interactive functionalities to filter or focus on certain nodes. Other more advanced functions are provided, e.g. a tool to find hidden patterns within nodes. Moreover, the framework has high scalability with respect to KG size.  \\

LodLive \cite{LodLive} is a software specifically developed for RDF graph visualization with SPARQL query interface. The user can search for an entity of interest from a specific graph database specifying the URI of the resource or a keyword representing the concept. Results are shown first displaying the target entity with a series of tools that allow to explore the connections, the description of the target entity and the connected entities. Different colors are used to distinguish classes. The software is open-source and can be integrated and personalized with plugins by developers, ensuring modularity; information about performances are not available; the interface is provided with a guide to use the available tools so it is quite intuitive, but may not be easy to use for non-expert users; query support goal is reached as the user does not need to learn and use SPARQL.

TigerGraph \cite{TigerGraph} is a company that developed a set of products for graph analytics. Among all their applications, there are also Knowledge Graph analysis and visualization. GraphStudio is one of their software tools aimed at gathering graph analysis and visual analytics into one easy-to-use graphical user interface. The user can search entities of interest using input keywords and also filter conditions with intuitive tools. There are many functionalities aimed to explore the graph: expansions of connections within nodes, finding paths or connections between nodes, edit data, and so on. The appearance of the resulting visualization can be changed for example with colors, sizes, displayed information about nodes. This can be done also with complex filters allowing the user to personalize the visualization making it clearer and insightful as possible. Despite being closed-source, the product is also modular as it is possible to directly code additional functionalities. More expert users can learn GSQL, an SQL-syntax like language that helps to write more complex queries. TigerGraph Insight is another product for graph visualization and analytics to find more valuable insights with no coding required. Aside all the classic visual analytics functionalities (similar to GraphStudio), there is also the possibility to create intuitive dashboards with other charts representing graph data.

EDUVIS \cite{sun2016eduvis} is an open-source visualization tool developed to effectively analyze educational data sparse on the Web. After gathering these data on a domain-specific (educational) KG, visualization techniques are applied for a better analysis and interaction. KG data are displayed in a network form as it is effective to highlight entities and relationships between them. Several colors and shapes of lines are employed to differentiate groups of entities and levels of importance of relationships, respectively. Additionally, the focus of the work is to acquire and visualize time information in the KG, building a event network based on timeline where event communities are positioned according to time axis. The user is provided with different interactive functionalities such as the possibility to track the click history on the graph and re-position the entities according to the clicked events. The developed framework seems to have an intuitive UI but it lacks the other aspects. Even if the work focuses on a domain-specific KG, the authors underlined that the framework can be generalized to other types of KGs.

ALOHA \cite{ALOHA} is a KG Visualization framework that aims at creating an intuitive and clear visualization of a dietary supplement KG and has been evaluated through a focus group experiment. The system gives query support as the user can easily specify data of interest with semantic queries and then the underlying architecture will compute it in the query language of the graph. Results are displayed with a graph containing requested information, so no other support charts are provided. However, the interface is provided with many functionalities to help the user navigating the KG and with additional information about data. Also colors are used to differentiate the semantic of concepts. Results of the focus group showed that improvements are necessary like the possibility to group similar nodes or to make textual information more visible.

CEPV \cite{CEPV} is a framework for the analysis and visualization of big knowledge graphs that provides useful and insightful visualizations for the users. The used approaches allow to efficiently extract data from a big graph database according to user's requirements, then a data processing step is executed to organize retrieved objects with their relationships with tree structures and finally visualizations are displayed with a subtree model or a textual visualization. The framework primarily reaches the performance goal as it has been created to solve the problem of querying big KGs. The query process is simplified because user's requirements are converted into query statements. The intuitiveness of UI is only partially achieved because, on the one hand, results can be displayed in text which the authors claim to be easier to understand for non-expert users, but on the other no other charts, recommendations or additional tools have been implemented.

The framework of Nararatwong et al. \cite{kgvisualization2020} is a web-based visualizer with a modular architecture allowing developers to easily add, extend, modify and delete functionalities. The system is composed by (i) a back-end subsystem, characterized by centralized access control, database connectors and an API module with different packages like CUI (Core User Interface) that allows to communicate with database connectors and to use common UI functionalities, and (ii) a front-end subsystem with a visualizer module that communicates with the back-end. The authors claim that the framework's interface is intuitive, thanks to a simple way to navigate the KG by searching for an entity of interest and retrieving, for example, all the neighbors with connections from the target entity. The user is also capable to customize the visualization with rules in order to better understand the context of data. Different basic tools are provided, but third-party plugins can be added for many kind of functions like additional charts or to make the framework more domain-specific. Also good performances are achieved as for the visualization of big KGs.

\begin{figure}[ht]
	\centering
    \includegraphics[angle=0,origin=c,width=0.9\textwidth]{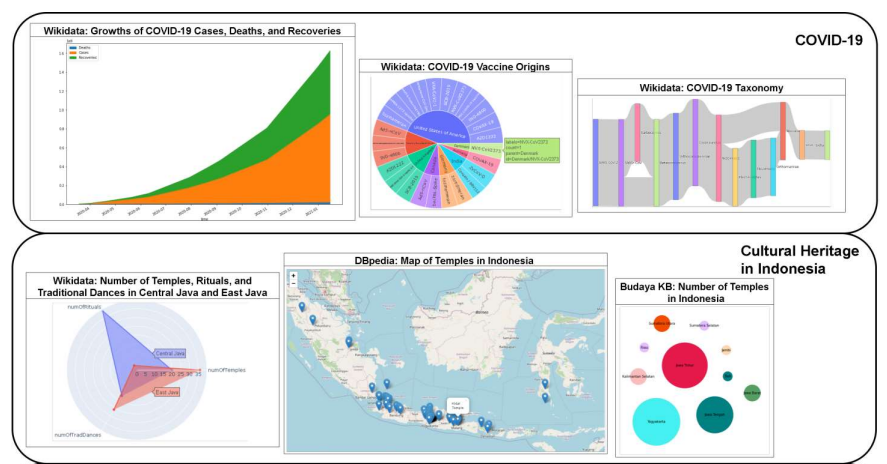}
    \caption{Examples of recommended chart for a KG - VizKG \cite{VizKG}}
    \label{fig:vizkg}
\end{figure}

VizKG \cite{VizKG} is a framework for the visualization of SPARQL results on KGs. A preprocessing stage is performed in order to parse and validate the query string in input and then results are given. To represent results of a SPARQL query in a more intuitive way, \textit{visualization recommendations} are computed with mapping rules and related charts are generated. So the results of a SPARQL query are mapped into comprehensive dashboards with charts like bar chart, pie chart, geographical maps, tree maps according to the input data types. In this framework, the intuitiveness of the visualization is reached thanks to the creation of recommended charts from the results. This also leads to facilitate the understanding of results for the user without the need to get insights from the graph structure and to make the querying process easier. According to authors, also modularity is accomplished with reference to chart implementations.
In Figure \ref{fig:vizkg} examples of recommended charts for KGs are shown.

Lim et al. \cite{KGcovid} proposed a visual analytics approach for Knowledge Graph containing information about COVID-19 papers. The work have a research-purpose so it has not the goal to create a framework for non-expert users. The goal is to provide support for predictive analyses, but the work still provides interesting ideas for intuitive visualizations. The displayed diagram is based on a force-directed graph drawing algorithm developed in a way in which nodes with a strong relationship attract each other and other nodes repel each other. Colors and shapes are used to differentiate the represented data, e.g. the thickness of the edges indicates the number of connections. The authors also performed statistical analysis on the graph related to the understanding of the structure using different types of diagrams, like box-plot and bar charts.

In KGNav framework \cite{wang2023kgnav} the main objective is to facilitate the user's understanding of KGs and the query process. In particular, the developed system allows a navigational interactive querying by (i) first building a Knowledge Graph Schema (KGS), i.e. a condensed KG which the user can easily understand, and then (ii) offering an intuitive user interface with a set of operators that enable the user to query the KG data in an easier way. The authors have demonstrated in practice the evaluation of framework's usability and intuitiveness by recruiting non-expert users to execute some query tasks and results led to describe the system with five main dimensions that are usability, professionalism, interactivity, practicality and learnability. The framework is based on a previous work of the authors \cite{liu2022kgvql} in which a novel visual query language called KGVQL is developed.

\section{Visualization with Knowledge Graph Embeddings}\label{sec:KGE_visualization}

Knowledge Graph Embeddings (KGEs) have proved helpful in the \textit{exploration} of KGs, but, as they are often produced by black-box models, they may be hard to manipulate in an \textit{explanation} perspective. For this reason \textcite{ettorre2022stunning} argue for joint visualization of KGs and KGEs as a way to benefit from both kinds of data, vector data supporting the exploration of KGs and network data supporting the explanation of KGEs.

In this section we review the techniques adopted for the visualization of KGEs, taking into account both perspectives and classifying the works that prioritize the exploratory or the explanatory one. A summary is provided in Table \ref{table:kgeviz}. A feature shared by all approaches is the representation of embedding vectors with some dimension reduction technique, such as PCA, t-SNE, or UMAP, in order to map the embeddings onto a 2-dimensional space. The plain visualization of the embeddings is usually accompanied by other visual elements aimed at enriching the analysis. Graph embeddings are generally computed before visualization, thus they do not affect the performance, except for dimension reduction techniques which are applied online. Thus methods in this section typically respect the performance requirement. Despite generally providing intuitive UIs, none of them implements modularity, i.e. allows for inclusion of new or modified views. Most of them allow for the use of multiple embedding models, either simultaneously, for comparison purposes, or alternatively. Finally, systems with an exploratory perspective mostly include query support, while explanatory-oriented ones don't, as query is not their main focus.

\begin{table}
    \begin{tabularx}{\textwidth} { 
  | >{\raggedright\arraybackslash}X 
  | >{\raggedright\arraybackslash}X |
  >{\centering\arraybackslash}X |
  >{\centering\arraybackslash}X |
  >{\centering\arraybackslash}X |
  >{\centering\arraybackslash}X |}
    \hline
    \textbf{Visualization framework} & \textbf{Perspective} & \textbf{Modularity} & \textbf{Intuitive UI} & \textbf{Performance} & \textbf{Query Support} \\
    \hline
    Making Sense of Search \cite{larson2020making} & Exploratory & & \checkmark & & \checkmark \\
    \hline
    VISION-KG \cite{wei2020vision} & Exploratory & & \checkmark & \checkmark & \checkmark \\
    \hline
    The Missing Path \cite{destandau2021missing} & Exploratory & & & &  \\
    \hline
    Stunning Doodle \cite{ettorre2022stunning} & Hybrid & & \checkmark & \checkmark & \checkmark \\
    \hline
    CorGIE \cite{liu2022visualizing} & Explanatory & & \checkmark & \checkmark &  \\
    \hline
    GEMVis \cite{chen2022gemvis} & Explanatory & & \checkmark & \checkmark &  \\
    \hline
    BiaScope \cite{rissaki2022biascope} & Explanatory & & \checkmark & \checkmark &  \\
    \hline
    KGScope \cite{yuan2024kgscope} & Exploratory & & \checkmark & & \checkmark \\ 
    \hline
  \end{tabularx}
  \caption{Summary of KG embedding visualization approaches, ordered by publication year.}
  \label{table:kgeviz}
\end{table}

\subsection{Exploratory perspective}

The exploratory approach to KG Embedding visualization consists in the use of embeddings to provide guidance in the exploration of a KG. This is often achieved by recommendation of entities and relations based on embedding vector similarity or clustering. Differences between approaches concern the type of embedding and the ancillary information displayed.

Making Sense of Search (\cite{larson2020making}) supports the exploration of navigation graphs on the Bing search engine by computing spectral embeddings (Laplacian Spectral Embedding, Adjacency Spectral Embedding), which help group nodes with a similar role in the graph. Their visualization is the only one to use a 3-dimensional projection of the node embeddings juxtaposed to a 2-dimensional node-link diagram, both displaying all nodes of the graph. Clusters can be computed either on the graph or on the vectors, and are represented in both visualizations using an equal color coding. Together with spatial visualizations, Making Sense of Search includes a \textit{hierarchy viewer}, i.e. a table that recommends entities from the same cluster as a certain query, sorted by their search frequency, and an \textit{embedding browser}, which uses \textsc{LineUp} visualization \cite{gratzl2013lineup} to rank entities based on vector similarity with the query, search frequency, and other attributes. 

VISION-KG \cite{wei2020vision} addresses the problem of summarizing the KG based on a queried entity in order to represent only a portion of the original graph. It exploits a combination of different (semantic and structural) embeddings to rank nodes that are linked to the query through a path, thus building and displaying a summarized graph centered on the query. Juxtaposed to it is another graph view: a \textit{topic-centric summarized subgraph}, obtained from the former by clustering embedding vectors. The user can thus select and alternatively visualize different entity clusters related to the same query. In both visualizations, color coding is used to indicate the type of each entity. Authors also use a rectangular shape to represent groups of entities that share the exact same relations as single nodes. This differentiates them from single entities, represented as circular nodes, and helps further summarize the graph.

The Missing Path \cite{destandau2021missing} adopts a different kind of embedding technique. It computes a sparse, high-dimensional, binary embedding of KG entities where each entry indicates whether the entity is not the root of a certain relation path, i.e. whether the entity does not have a certain property or a chain of properties. Here the specific purpose of the visualization is indeed to analyse the incompleteness in KGs. Entities are then plotted in a 2-dimensional space via dimension reduction. The map is combined with a mirrored histogram, which provides an aggregated view of the completeness of each path (i.e. the percentage of entities that have that path) in the full KG and in a subset selected by the user. The mirrored histogram allows for direct comparison between the full set of entities and the subset. Moreover, color coding is applied to path labels to signal which paths are missing from the selected subset of entities, and which paths are specific to that subset.

Stunning Doodle \cite{ettorre2022stunning} is the only tool that explicitly integrates an exploratory and an explanatory perspective. It provides a simple 2-dimensional node-link diagram centered on a selected entity and displaying all entities in its one-hop neighborhood. Each entity node may then be expanded so to include a portion of the two-hop neighborhood of the central entity without making the visualization unmanageable. Entity embeddings, obtained by any machine learning model, are used to compute the distance from the central entity. Such distance is encoded in color luminance, which therefore provides guidance to the user by indicating entities that are similar to the selected one.

KGScope \cite{yuan2024kgscope} adopts TransR \cite{lin2015learning} embedding model, which maps entities to as many vector spaces as the number of relations in the KG. Here the node-link diagram, plotted with force-directed layout, is juxtaposed to one or more \textit{embedding views}, i.e. 2-dimensional scatterplots representing the embedding vectors of entities in the graph. Since TransR produces different embeddings for different relations, each of these plots refers to a specific relation. Node shape indicates the type of entity with respect to the relation (head, tail, or both), while node luminance and saturation encode a metric called \textit{peculiarity}, which is aimed at measuring the unexpected information captured by entity embeddings. This metric provides guidance to explore entities of interest, whose links are displayed upon selection. Finally, clustering embeddings proved successful in automatically distinguishing categories of entities (e.g. person, city, etc), also encoded in the node-link diagram by different colors.

\subsection{Explanatory perspective}

In \cite{li2018embeddingvis}, the authors point out the reasons for an explanatory approach to graph embeddings. They underline 3 challenges that can be addressed by graph embedding visualization: (1) the analysis of the embedding latent meaning (abstract representation), (2) the exploration of the embedding space, and (3) the fine-grained and comparative analysis of the semantic and structural information retained by embedding models. Though not specific to KGs, the relevance of their approach lies in the identification of 3 levels of visual analysis: the \textit{cluster-level}, which explores clusters of similar nodes, the \textit{instance-level}, which compares node neighborhoods and embedding similarities, and finally the \textit{structure-level}, which analyzes the distributions of node-neighborhood average embedding distances.

While several tools exist for the analysis of embeddings, including graph embeddings, Stunning Doodle \cite{ettorre2022stunning} is the only tool that addresses the problem from the specific perspective of KGs. It does so by integrating graph visualization with embedding-based information, i.e. distance from a queried entity encoded in the node luminance. Such an approach addresses only the \textit{instance-level}, as it allows for comparison between embeddings of adjacent nodes.

Most approaches for explanation of graph embeddings have been developed for general graphs. Despite not being tailored to KGs, we still deem these methods remarkable for their generality and their capacity to address the points highlighted in \cite{li2018embeddingvis}, which make them applicable in our context as well.

\begin{figure}
    \centering
    \begin{subfigure}{0.4\textwidth}
        \includegraphics[width=\textwidth]{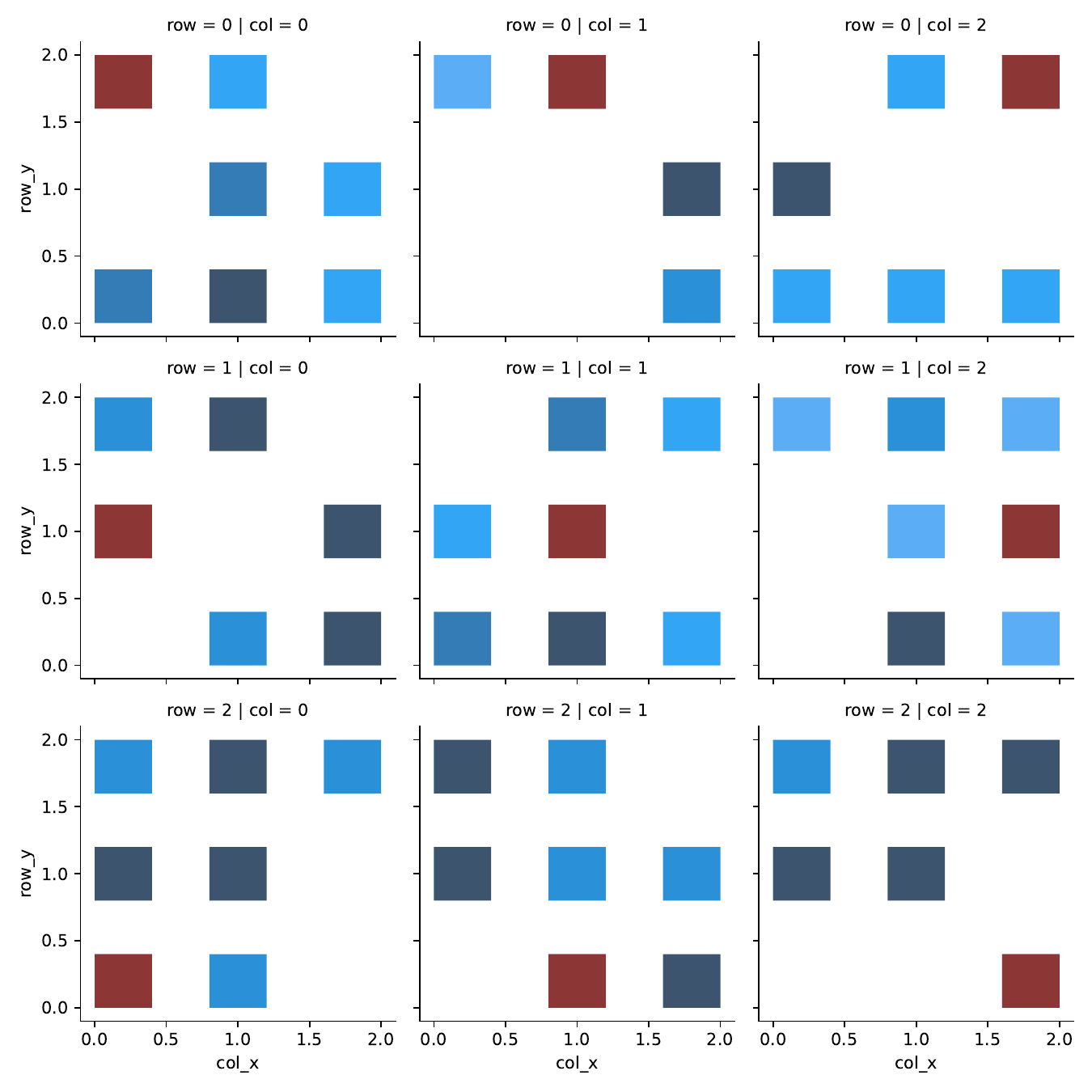}
        \caption{Latent neighbor block view.}
        \label{fig:od}
    \end{subfigure}
    \hfill
    \begin{subfigure}{0.4\textwidth}
        \includegraphics[width=\textwidth]{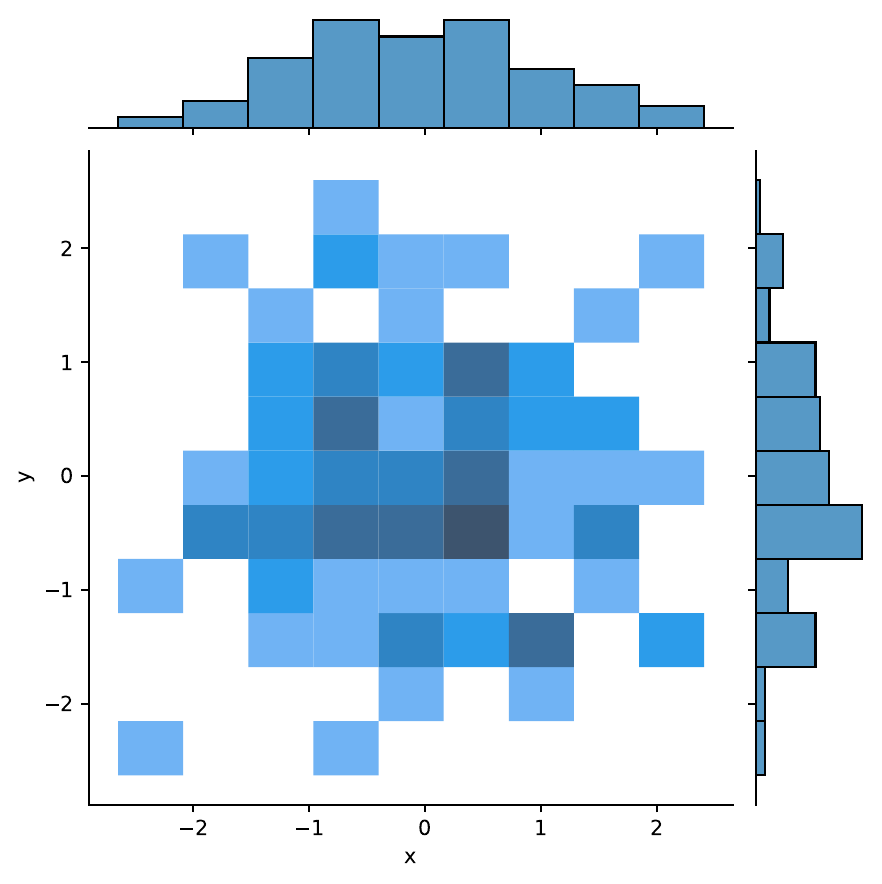}
        \caption{Distance comparison view.}
        \label{fig:jointplot}
    \end{subfigure}
    \caption{Toy examples of CorGIE latent neighbor block view with 9 blocks (a) and distance comparison view (b).}
    \label{fig:corgie}
\end{figure}

Though developed for embeddings produced by Graph Neural Networks, CorGIE (\cite{liu2022visualizing}) accomodates the analysis of any kind of graph embeddings. The tool is based on the visualization and comparison of graph nodes in 3 spaces: the graph topological space, the node feature space, and the latent embedding space. For this purpose, it presents 6 simultaneous views. One view represents the nodes in a 2-dimensional reduction of the latent space, using background coloring rather than node coloring to highlight cluster areas. The node feature space is depicted in another view as a histogram matrix, each histogram representing one node feature. Finally, two views deal with the topological space, representing (1) the whole node-link diagram with a force-directed layout, and (2) the connections in 1-hop and 2-hop neighborhoods of one or more focal groups of nodes selected by the user. 
\textit{Instance-level} analysis is provided by the \textit{latent neighbor block} view (see Figure \ref{fig:od} for an example), which draws from Origin-Destination maps in geo-spatial networks to partition the latent space in 64 blocks and show, in each one, the 2-dimensional spatial distribution of neighbors of nodes in that block. This approach allows to draw a correspondence between distances in the topological space and in the latent space, to verify whether the embedding model preserves the structure of the graph. \textit{Structure-level} analysis is conducted in the \textit{distance comparison} view (see Figure \ref{fig:jointplot} for an example), which represents the joint distribution of node-node distances in the latent space and in one of the other spaces (topological or node feature). This view allows to derive correlations between distances in different spaces. Finally, \textit{cluster-level} analysis is enabled by giving the user the possibility to select focal groups of nodes, which are then depicted in a different color in the latent space view, and in separate plots in the feature space and \textit{distance comparison} views.

In GEMVis (\cite{chen2022gemvis}), multiple Graph Embedding models can be analyzed and compared in a5 view framework. Besides plotting the node-link diagram in force-directed layout and the 2-dimensional scatterplot of node embeddings, it depicts the distribution of node centrality measures on a parallel coordinate chart, which makes the distributions immediately comparable. In these 3 views, the user is given the possibility to select a subset of nodes the will be highlighted by turning the other nodes grey. A red-green gradient color palette is also made available to better deal with cases in which the number of selected nodes is large. Comparison between embedding models is implemented in the \textit{R\_node metric} view, which represents in a radar plot the capacity of each selected embedding model to retain structural information from the original graph. This capacity is captured in 9 $R^2$ coefficients obtained by performing a nonlinear regression of the node centrality metrics on the node embeddings. This allows for quick comparison between different models, which are represented by lines of different colors in the plot. Finally, the tool also includes a historical record view, allowing to retrieve the statistics of models evaluated in the past.

BiaScope (\cite{rissaki2022biascope}) focuses specifically on unfairness in Graph Embedding models. In order to measure the unfairness of a node embedding, they compute its average Euclidean distance with respect to its neighbors, then normalizing the scores first with respect to the node degree and finally in $[0;1]$. For the qualitative analysis of unfairness, the system includes 3 views. The first one provides a statistical summary on network characteristics, such as number of nodes, edge density and average clustering coefficient, together with a histogram of node degrees. The second view allows to compare two node embedding models by juxtaposition of two node-link diagrams where precomputed node unfairness is encoded in node color, which lays on a gradient from white to red. Thus, the system makes it possible to immediately recognize which nodes, or groups of nodes, contribute to the unfairness of an embedding model. The third view, finally, focuses on one node, upon selection by part of the user. Specifically, it juxtaposes the 1-hop ego network in the node-link diagram and the 2-dimensional projection of the embeddings of the nodes in such subgraph. Here red color, together with node size, is used to convey a popout effect for the focal node. This view helps diagnose the causes of unfairness: the further away a node is mapped in the embedding space, the higher its contribution to the focal node unfairness. Despite this diagnose view, the system doesn't provide any other interactivity feature, for instance semantic zooming or details-on-demand. 

Despite being applicable to KGs as a particular form of a general graph, the presented approaches are all limited to node (entity) embeddings and do not deal with relation embeddings, which are part of the outputs of KGE models and constitute their peculiarity. Nevertheless, such limitation is shared by the approaches built for KGs as well, thus it indicates one possible future research direction.

\section{Concluding Remarks}\label{sec:concl}

In this work we have reviewed the state-of-the-art visualization techniques for Knowledge Graphs and Knowledge Graph Embeddings. Having identified the main challenges that a visualization system has to address, namely \textit{modularity, intuitive UI, performance}, and \textit{query support}, we have analyzed the main techniques that aim to solve them, with a special attention dedicated to the integration of graph embedding techniques into the visualization system. In particular, we have categorized such techniques in two classes: those that use embeddings in an \textit{exploratory} perspective and those that take an \textit{explanatory} perspective.\\
We found that most methods include the visualization of the graph, often in a simplified form, as a node-link diagram, accompanied by other charts that aim at representing statistics. The diagram usually has a force-directed layout, which helps prevent the overlapping of nodes and edges. Node selection, filtering and brushing functionalities are also widely adopted to facilitate visual analysis. Other functionalities, such as semantic zooming, are rarely supported, which makes them possible directions for future work. In the context of KGs, semantic zooming in particular poses several research questions, such as which entities to hide when zooming out.\\
In line with \cite{munzner2014visualization}, the most widely adopted visual channel to represent classes and clusters of entities is color hue. The frameworks that do not highlight classes or clusters tend to use color for popout effect of selected nodes or as a gradient to map nodes to a certain metric. Other charts include scatterplots, which are effective for mapping 2-dimensional projections of the node embeddings, and histograms, for statistics. Statistics, however, are always computed on entities, while relations are generally overlooked.\\
All in all, the intuitive UI and performance requirements are usually met, in the case of embeddings thanks to offline computation. Query support is provided in general KG visualization frameworks as well as in frameworks that use KG embeddings in an exploratory perspective, but not in those that take an explanatory perspective on them. Finally, frameworks are rarely modular, so the visualizations are typically fixed.\\
As a consequence, modularity is a promising direction of future research, together with the visualization of relations. These would stand out as peculiarities of KG visualization against general graph visualization. Adding query support to the KGE explanatory frameworks would also add potential to the analysis of KGEs at the instance level. Finally, exploring new functionalities will likely pose new challenges to be addressed with the purpose of knowledge representation in mind.

\section*{Acknowledgments}

This publication is part of the project PNRR-NGEU which has received funding from the MUR – DM 118/2023.

\printbibliography

\end{document}